\documentclass[traditabstract]{aa} 
\usepackage{graphicx,longtable,cancel}
\usepackage{txfonts,multirow}
\begin{document}
\bibliographystyle{aa}
\newcommand{\um}{$\mu$m}
\newcommand{\uJy}{$\mu$Jy}
\newcommand{\ergs}{{erg\,s$^{-1}$}}
\newcommand{\ergscm}{{erg\,s$^{-1}$\,cm$^{-2}$}}
\newcommand{\kms}{{km\,s$^{-1}$}}
\newcommand{\msun}{${\rm M}_{\odot}$}
\newcommand{\lsun}{${\rm L}_{\odot}$}
\newcommand{\msunyr}{${\rm M}_{\odot}~{\rm yr}^{-1}$}
\newcommand{\lir}{$L_{\rm IR}$}
\newcommand{\Lbol}{$L_{\rm bol}$}
\newcommand{\ms}{$M_{\rm star}$}
\newcommand{\mg}{$M_{\rm gas}$}
\newcommand{\mh}{$M_{\rm halo}$}
\newcommand{\mbh}{$M_{\rm BH}$}
\newcommand{\mui}{$\mu^{-1}$}
\newcommand{\cohh}{CO$\to$H$_2$}
\def\Zsun{Z_\odot}
\def\gtsima{$\; \buildrel > \over \sim \;$} 
\def\ltsima{$\; \buildrel < \over \sim \;$} \def\prosima{$\; \buildrel \propto \over \sim \;$} 
\def\gsim{\lower.5ex\hbox{\gtsima}}
\def\simgt{\lower.5ex\hbox{\gtsima}} 
\def\simlt{\lower.5ex\hbox{\ltsima}} 
\def\lsim{\lower.5ex\hbox{\ltsima}}
\def\msole{M_\odot}
\def\rsole{R_\odot}
\def\msun{M_\odot}
\def\rsun{R_\odot}
\def\mdot{\dot{M}}
\def\OVI{\hbox{O~$\scriptstyle\rm VI$}}
\def\invMpch {~h~{\rm Mpc}^{-1}}
\def\Mpch {~h^{-1}~{\rm Mpc}}
\def\kpch {~h^{-1}~{\rm kpc}}
\def\swift {{\it Swift}}
\def\xmm {{\it XMM-Newton}}
\def\chandra {{\it Chandra}}
\makeatletter
\def\@biblabel#1{\hspace*{-\labelsep}}
\makeatother

\title{A physical scenario for the high and low X--ray luminosity states in the transitional pulsar PSR J1023+0038}

\author{
S. Campana\inst{1}, 
F. Coti Zelati\inst{2,1,3},
A. Papitto\inst{4,5},
N. Rea\inst{3,4},
D. F. Torres\inst{4,6},
M. C. Baglio\inst{2,1},
P. D'Avanzo\inst{1}
}
\institute{INAF, Osservatorio Astronomico di Brera, Via E. Bianchi 46, I-23807, Merate (LC), Italy\\
 \email{sergio.campana@brera.inaf.it}
\and
Universit\`a dell'Insubria, Dipartimento di Scienza e Alta Tecnologia, Via Valleggio 11, I-22100 Como, Italy 
\and
Anton Pannekoek Institute for Astronomy, University of Amsterdam, Postbus 94249, NL-1090-GE Amsterdam, The Netherlands
\and
Instituto de Ciencias de l'Espacio (ICE, CSIC-IEEC), Campus UAB, Carrer Can Magrans s/n, E-08193 Barcelona, Spain
\and
INAF, Osservatorio Astronomico di Roma, Via di Frascati 33, I-00044, Monteporzio Catone (Roma), Italy
\and
ICREA Instituci\'o Catalana de Recerca i Estudis Avan\c{c}ats, E-08010 Barcelona, Spain}

\date{}

\abstract{
PSR J1023+0038 (J1023) is a binary system hosting a neutron star and a low mass companion. J1023 is the best studied transitional
pulsar, alternating a faint eclipsing millisecond radio pulsar state to a brighter X--ray active state. At variance with other Low Mass X--ray 
binaries, this active state reaches luminosities of only $\sim 10^{34}$ erg s$^{-1}$, showing strong, fast variability. In the active state, 
J1023 displays: $i)$ a high state ($L_X\sim 7\times 10^{33}$ erg s$^{-1}$, 0.3--80 keV) occurring $\sim 80\%$ of the time and during 
which X--ray pulsations at the neutron star spin period are detected (pulsed fraction $\sim 8\%$); $ii)$ a low state ($L_X\sim 
10^{33}$ erg s$^{-1}$) during which pulsations are not detected ($\lsim 3\%$); and $iii)$ a flaring state during which sporadic 
flares occur in excess of $\sim  10^{34}$ erg s$^{-1}$, with no pulsation too. The transition between the high and the low states 
is very rapid, on a $\sim 10$ s timescale. Here we put forward a plausible physical interpretation of the high and low states based 
on the (fast) transition among the propeller state and the radio pulsar state. We modelled the XMM-Newton spectra of the high, 
low and radio pulsar states, finding a good agreement with this physical picture. }

\authorrunning{Campana et al.}

\maketitle

\section{Introduction}

Millisecond radio pulsars (MSPs) are the final outcome of the accretion process taking place in Low Mass X--ray Binaries (LMXBs). 
The recycling scenario has been suggested to explain the spin up of a neutron star accreting from a disc in a LMXB to millisecond 
spin periods resulting, in the end, in the turn on of the neutron star as a MSP  (Alpar et al. 1982; Smarr \& Blandford 1976).
Confirmation of this scenario came with the discovery of coherent periodicities during type I X--ray bursts in persistent LMXBs 
(Strohmayer et al. 1996) and, soon after, with the detection of coherent pulsations in the X--ray flux of the transient LMXB 
SAX J1808.4--3658 (Wijnands \& van der Klis 1998). 

At the end of the evolutionary path, ``spider" MSPs were discovered before X--ray periodicities in LMXBs 
(e.g. Fruchter, Stinebring \& Taylor 1988).
These are binary systems consisting of a radio MSP and a low mass companion (either a white dwarf, black widow, or a low 
mass star, redback), showing signs of interaction between the radio pulsar and the companion by means of extended radio eclipses, 
originating from spatial regions larger than the companion star, e.g. in an intra-binary shock front (Roberts 2011).
In the past it was not clear on which timescale the transition from LMXB to MSP occurs and if it happens only once or repeatedly.
Recently, systems alternating between the MSP and the LMXB status were found.
First, strong indirect evidence that the latter was indeed the case was put forward by PSR J1023+0038 (Archibald et al. 2009, 
see also XSS J12270--4859, De Martino et al. 2010; Bassa et al. 2014). These systems were classified as magnetic cataclysmic 
variables (i.e. accreting systems) and later on revealed as radio pulsars. 
The direct transition between the two states was caught in IGR J18245--2452 (in the globular cluster M28, Papitto et al. 2013),
where an X--ray source coincident with a radio pulsar went in outburst (showing X--ray pulsations at the spin period of the radio 
pulsar) and turned to quiescence reappearing again as a radio pulsar, within a few weeks.
These sources are now known as transitional pulsars.

PSR J1023+0038 (J1023 in the following) is a particularly well-studied system due to its proximity ($1.37\pm0.04$ kpc, 
Deller et al. 2012) and brightness.
J1023 underwent a state change from MSP to a brighter (accreting) state during June 2013. This implied the 
disappearance of the radio pulsed signal (Stappers et al. 2014) as well as an increase by  a factor of $\sim 5$ in the 
GeV flux (Stappers et al. 2014; Takata et al. 2014). In the 0.5--10 keV band the system brightened by a factor of $\sim 30-300$, 
reaching luminosities up to a few $10^{34}$ erg s$^{-1}$ (Patruno et al. 2014; Takata et al. 2014; Coti Zelati et al. 2014; 
Bogdanov et al. 2015) with respect to the quiescent (radio pulsar) state 
(Archibald et al. 2010; Bogdanov et al. 2011). 
In the optical, the system brightened too by $\sim  1$ mag and showed also the presence of several broad 
emission lines typical of an accretion disc (Halpern et al. 2013; Coti Zelati et al. 2014).

Deep X--ray observations of the active state have been carried out. Unlike transient LMXBs, J1023 as well as the other 
transitional pulsars shows strong X--ray variability (e.g. Linares 2014). J1023 varies its X--ray luminosity over the 
$\sim 10^{32.5}-10^{34.5}$ erg s$^{-1}$ range showing three distinct states: 
$i$) a high luminosity state ($\sim 7\times 10^{33}$ erg s$^{-1}$, 0.3--79 keV Tendulkar et al. 2014) occurring for the majority 
of the time ($\sim 80\%$) and during which coherent pulsations are observed with a rms pulsed fraction of $\sim 8\%$; 
$ii$) a low luminosity state ($20\%$ of the time) with a lower X--ray luminosity $\sim 10^{33}$ erg s$^{-1}$ (0.3--79 keV) 
and during which no pulsations are detected (rms $95\%$ upper limit of $2.4\%$); $iii$) a flaring state with higher luminosity 
up to a few $10^{34}$ erg s$^{-1}$, with no pulsations, occurring for a few percent of the time (see Archibald et al. 2015; 
Bogdanov et al. 2015).
Transitions from and to the low and high states are very rapid with a timescale of the order of $\sim 10$ s. 
Similar variability has also been observed in the optical band showing ingress and egress times in the 12--35 s range 
with 20 s median (Shahbaz et al. 2015). 
This fast time variability among a high and a low X--ray state has also been observed in XSS J12270--4859 
(Bogdanov et al. 2014; Papitto et al. 2015).

When in quiescence J1023 is detected as a radio pulsar and shows a smaller X--ray luminosity of $\sim 9\times10^{31}$  
erg s$^{-1}$ (0.5--10 keV), as well as an X--ray pulsed fraction of $\sim 11\pm2\%$ (Archibald et al. 2010; Bogdanov et al. 2011).

\begin{table*}[!thb]
\caption{J1023 XMM-Newton/Chandra observation log.}
\begin{center}
\begin{tabular}{*6c}
\hline
\hline
Obs. ID.          & Obs. start     &Instrument  & Duration& High state        & Low state    \\ 
                        &(YYYY-MM-DD)&                &  (ks)       &  dur. - ks (counts)     & dur. - ks (counts) \\              
\hline
0720030101 & 2013-11-10 & pn (FT)       & 128.4     & 85.2 (287431) & 25.0 (21482)\\
                        &                        & MOS (SW) & 128.8     & 83.6 (169920) & 24.6 (8466)\\
\hline
0742610101 & 2014-06-10 & pn (FT)       & 116.7    & 63.8 (189226) & 26.2 (12453)\\
                         &                       & MOS (SW) & 118.5    & 62.7 (138779) & 25.9 (8379)\\
\hline                                                 
0748390101 & 2014-11-21 &pn (FT)        & 32.4& &\\
                         &                       & MOS (SW) & 34.3& &\\
0748390501 & 2014-11-23 &pn (FT)        & 32.9      &61.0 (180708) & 19.3 (9386)\\
                         &                       & MOS (SW) & 34.8      & 60.0 (128561)& 19.1 (6026)\\
0748390601 & 2014-11-28 &pn (FT)        & 16.9& &\\
                         &                       & MOS (SW) & 20.6& &\\
0748390701 & 2014-12-17 &pn (FT)        & 32.5& &\\
                         &                       & MOS (SW) & 34.4& &\\
\hline
\hline
0560180801 &2008-11-26 & MOS (FW) & 34.2 & \multicolumn{2}{c}{Quiescence (2136)} \\
11075  (Chandra)                 &2010-03-26& ACIS-S       & 86.2  & \multicolumn{2}{c}{Quiescence (3306)}\\ 
\hline
\hline
\end{tabular}
\end{center}
\noindent Exposure times for MOS refers to the MOS2 detector (MOS1 times are very similar), after cleaning for soft proton flares.\\
\noindent MOS counts for high and low state spectra are the sum of MOS1 and MOS2 counts.\\
\noindent Exposure times and total counts for the last set of observations in the transitional state were summed up.\\
\label{obslog}
\end{table*}

A few scenarios have been put forward trying to account for the wealth of data from the radio to the GeV energy range. 
These involve a propeller mechanism, an engulfed radio pulsar, direct accretion onto the neutron star, as well as a 
possible radio jet (Patruno et al. 2014; Takata et al. 2014; Papitto \& Torres 2015; Coti Zelati et al. 2014;  Li et al. 2014; 
Deller et al. 2015; to mention a few). 
However, a physical interpretation for the fast transitions mentioned above is not available at present. In this paper we focus on 
low and high X--ray emission states, and in particular, we focus on describing a possible mechanism that may generate these transitions.
We base our analysis on XMM-Newton data, which provide the highest signal to noise data. In particular, we focus on one 
possible mechanism showing that it works for this source, without willing to  assess its unicity.

In Section 2 we briefly describe the physical interpretation of state changes. In Section 3 we describe the extraction 
of XMM-Newton spectra and in Section 4 their fitting. A brief timing analysis is described in Section 5. 
In Section 6 we discuss our results and draw conclusions.

\section{The physical picture}

The neutron star's parameters were determined when the system was hosting an active radio pulsar, with a spin period of 1.69 
ms and a dipolar magnetic field of $9.7\times 10^7$ G (Archibald et al. 2009; Deller et al. 2012). 
The binary system has an orbital period of 4.75 hr and the companion is a $\sim 0.24\,\msole$ star,
possibly not filling entirely its Roche lobe (McConnell et al. 2014).

The high X--ray state is characterised by X--ray pulsations (rms pulsed fraction $\sim 8\%$) and a high luminosity 
($\sim 7\times10^{33}$ erg s$^{-1}$, 0.3--79 keV, resulting in a bolometric correction $\sim 2$ with respect to the 
0.3--10 keV energy band, Papitto \& Torres 2015; Tendulkar et al. 2014). 
The most direct interpretation to account for coherent pulsations is that some matter reaches the neutron star surface. 
Taking at face value, if the high state luminosity is converted into an accretion rate ($\mdot$) as $L=G\,M\,\mdot/R$, 
with $M$ and $R$ the neutron star mass ($M=1.4\,\msole$) and radius ($R=10$ km), and $G$ the gravitational constant, 
one obtains $\mdot=4\times 10^{13}$ g s$^{-1}$.
With this rate one can compute the Alfv\'en radius $r_{\rm A}$ at which the pressure of the neutron star magnetosphere 
is able to halt the inflowing matter in spherical symmetry ($r_{\rm A}\sim \mu^4/(2\,G\,M\mdot^2)^{1/7}$, with $\mu=B\,R^3$ 
the magnetic dipole moment).
Here $\mdot$ is the effective mass inflow rate at the magnetospheric boundary. It can be lower than the mass inflow rate 
from the companion star if matter is lost during the inflowing process or viscously transmitted through the disc at a lower rate. 
Assuming the mass accretion rate estimated from the X--ray luminosity one gets $r_{\rm A}\sim 148$ km. 
The Alfv\'en radius in case of disc accretion differs by a small factor from the one derived in spherical symmetry,
$r_{\rm m}=k_{\rm m}\,r_{\rm A}$, with $k_{\rm m}$ ranging from 0.5 to 1 (Ghosh \& Lamb 1978; Bozzo et al. 2009; Papitto \& Torres 2015). 
With these boundaries the magnetospheric radius for disc accretion is $r_{\rm m}\sim 74-148$ km.

The corotation radius (at which matter in Keplerian orbit corotates with the neutron star, $r_{\rm c}=(G\,M\,P^2/4\,\pi^2)^{1/3}$, 
with $P$ the spin period) for J1023 is $r_{\rm c}=24$ km and the light cylinder radius (at which field lines anchored to the neutron 
star rotate at the speed of light, $r_{\rm lc}=c\,P/2\,\pi$, with $c$ the light speed) is $r_{\rm lc}=80$ km.
According to standard accretion theory (e.g. Illarionov \& Sunyaev 1975; Campana et al. 1998) if $r_{\rm m}>r_{\rm lc}$, 
as suggested above, matter cannot accrete onto the neutron star surface, rather the neutron star should get rid of the 
infalling matter and shine as a radio pulsar.\\
This simple calculation shows that we need more pressure from the infalling matter to push deeper inside the 
neutron star magnetosphere, in order to have accretion of some matter on the surface and generate X--ray pulsations.
The easiest way is  to have a larger amount of accreting matter (i.e. a larger $\mdot$ with respect to that estimated from 
the X--ray luminosity).  This matter, however, should not release its entire gravitational energy content 
even if stopped at the magnetospheric boundary (i.e. $G\,M\mdot/r_{\rm m}$), simply because we do not detect it.
This additional matter cannot evaporate along its accretion flow toward the neutron star nor can it be channelled 
into a large base jet, because we need matter pressure at the magnetospheric boundary. A possibility is to have an 
advection-dominated accretion flow where radiation is trapped into the disc and only a percentage of what is produced 
can be radiated (Narayan et al. 1996). 
With more matter pressure the innermost disc boundary can be pushed within the light cylinder radius and close to the 
corotation radius.
If this happens, matter can still no longer accrete unimpeded onto the neutron star surface (propeller regime), being 
halted by the super-Keplerian rotation of  the magnetosphere. In the propeller regime however, some matter can leak 
through the magnetosphere reaching the star surface  (e.g. Romanova et al. 2004, 2005; Ustyugova et al. 2006). 
This can generate the observed coherent pulsations in X--rays. 
If this were the case, in the X--ray band we should observe two contributions: radiation from the advective disc up to distance 
between the corotation and the light cylinder radius, as well as a pulsed X--ray component arising from matter leaking 
through the magnetosphere and accreting onto the surface.  
A non-thermal component associated to a hot corona above the disc, as usually observed in X--ray binaries and in 
active galactic nuclei, should complete the picture. More compelling to this case can be the action of the neutron star magnetosphere 
onto the infalling matter generating a non-thermal continuum as a result of the propeller action (Papitto \& Torres 2015).

\begin{figure}[!ht]
\hskip 1.truecm
\includegraphics[width=0.4\textwidth]{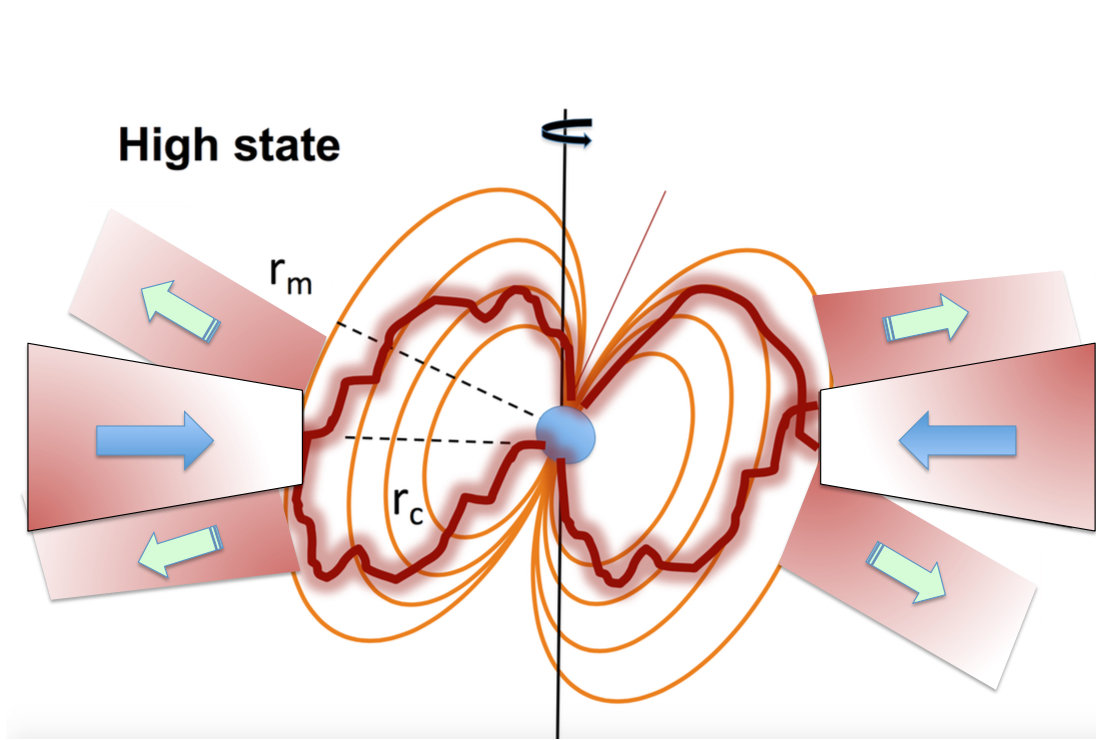}\\
\vskip 0.3 truecm
\includegraphics[width=0.5\textwidth]{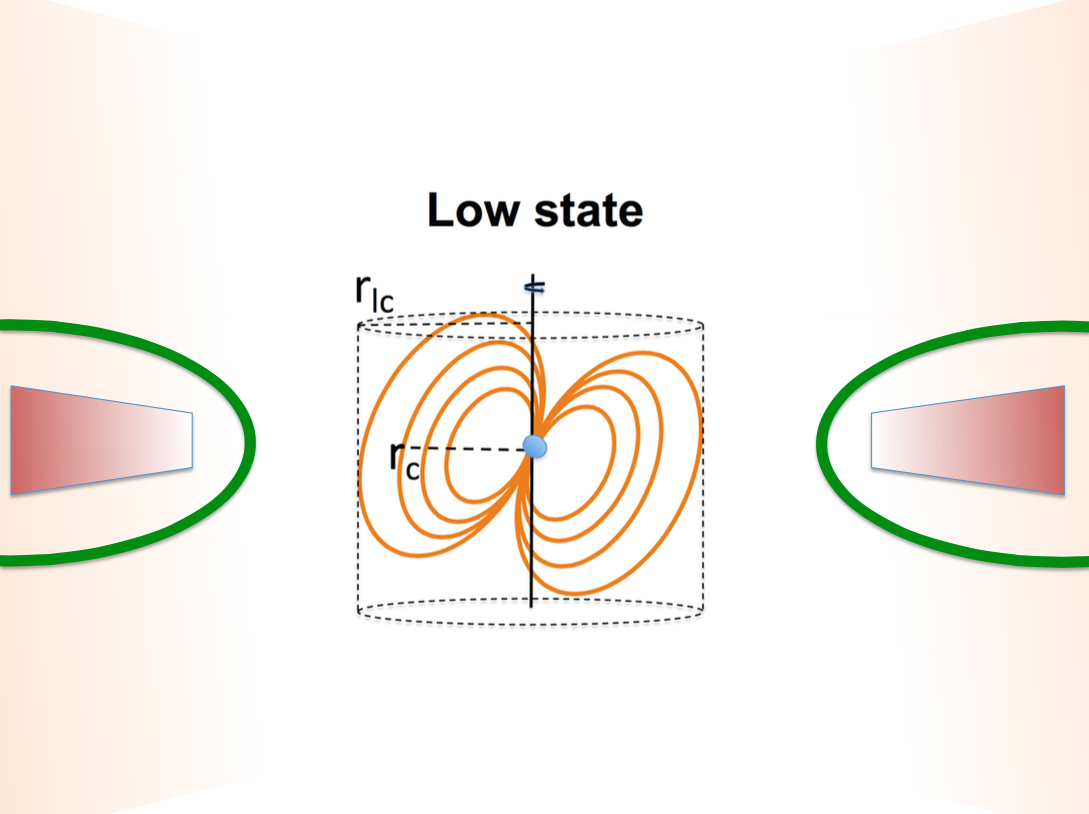}\\
\vskip 0.3 truecm \hskip 1.5truecm
\includegraphics[width=0.3\textwidth]{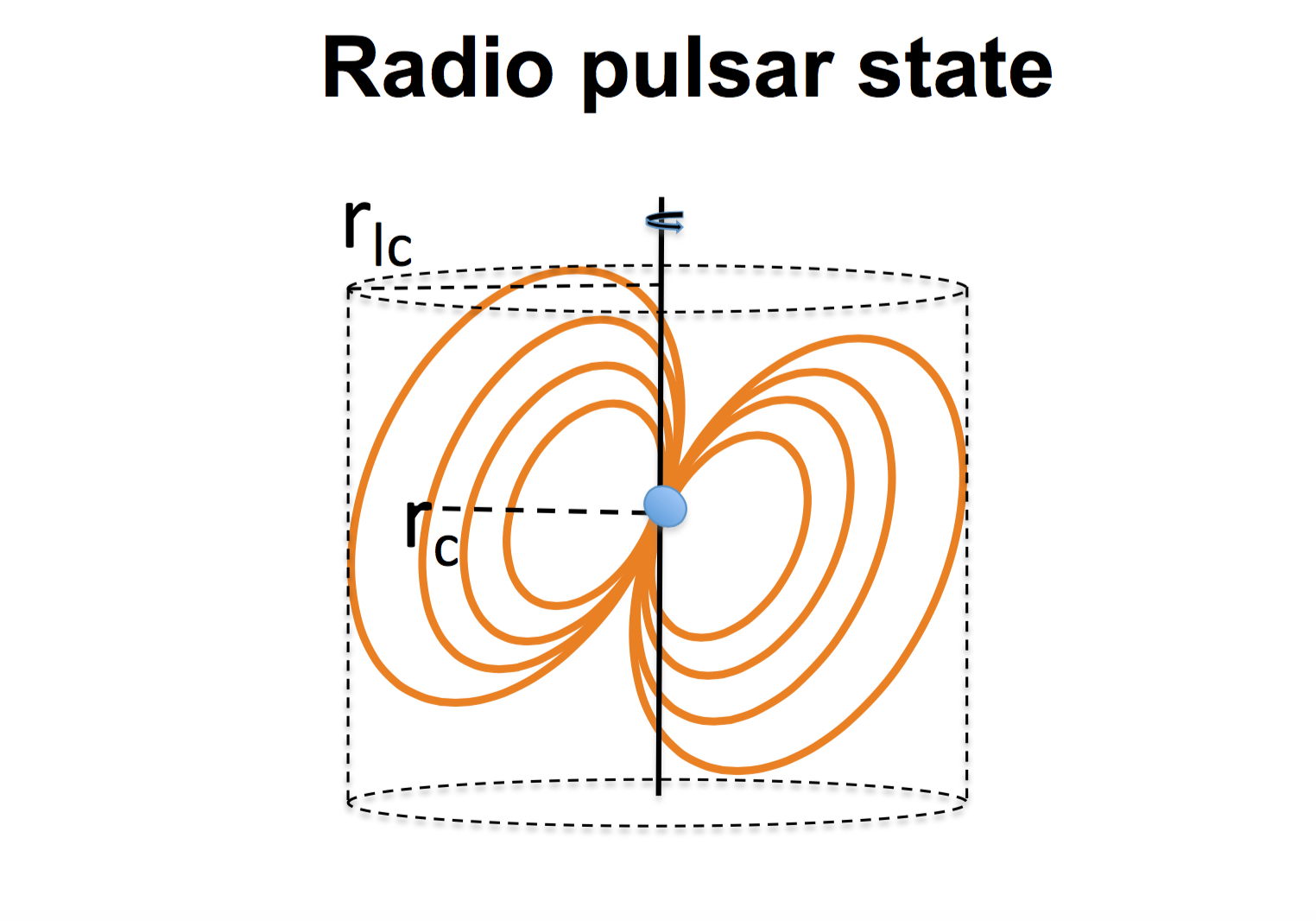}\\
\caption{Cartoon view (not to scale) of the three states of J1023. The high state is depicted in the top panel, the low state in the middle one and 
the pulsar, quiescent state in the lowest panel. The coronation radius is 24 km and the light cylinder radius is 80 km. 
The magnetospheric radius changes with the J0123 state. See text for more details.}
\label{fig1}
\end{figure}

The low X--ray state instead is characterised by the lack of X--ray pulsation, with an upper limit of $< 2.4\%$ on the rms pulsed 
fraction, well 
below the value observed in the high state. The 0.3--10 keV luminosity is a factor $\sim 6$ lower than the high state one, at a level of 
$\sim 5\times 10^{32}$ erg s$^{-1}$ (Archibald et al. 2015; Bogdanov et al. 2015), resulting in a 0.3--79 keV luminosity of $\sim 10^{33}$ 
erg s$^{-1}$ (Tendulkar et al. 2014). In order to have pulsations some matter has to reach the neutron star surface (unless one accepts
a rotationally powered pulsar to remain permanently active, which is subject to strong caveats, see Papitto \& Torres 2015).
According to numerical simulations, in the propeller regime some matter always finds its way to the star surface, possibly except 
when a `strong' propeller sets in (Romanova et al. 2005). 
However, due to the low neutron star magnetic field and fast spin period, the light cylinder radius is of the same order of the corotation 
radius and a strong propeller is difficult to build up (D'Angelo \& Spruit 2010).
We are then led to require that the system must expel the incoming matter outside the light  cylinder radius to avoid accretion.
Pushing out the material from the light cylinder will reactivate the pulsar particle wind and prevent matter from falling onto the neutron star. 
A radio pulsar can eventually turn on, even if its reactivation requires more stringent constraints. If this were the case, its detectability 
would be problematic due to the enshrouding by a large wealth of material. 
A shock front between the relativistic particle wind and the incoming matter should develop, as observed in several 
``spider'' pulsars (Tavani 1991; Kulkarni et al. 1992). This front can convert $\sim 1-10\%$ of the rotational power into X--rays with a 
typical synchrotron spectrum with $\Gamma \sim 2$ (Arons \& Tavani 1993).  
Solutions with a disc surviving just outside the light cylinder exist (Ek\c{s}i \& Alpar 2005). This outer disc should remain in place and
accumulate matter during the low state. Finally, the particle wind is quenched and disc advances back toward the corotation radius.\\
Even if matter is prevented from falling directly onto the neutron star surface, the polar caps will still be hot because of the short timescale 
of the state transitions, and can in principle still generate pulsations at a lower level.
The polar cap temperature when matter is not accreting is driven by the outer and inner crust heat content and it should be cooler 
(or hotter) than the temperature experimented during the accretion phase, depending on the temperature contrast among the innermost and 
outermost regions. The cooling time of the neutron star atmosphere is very short so that the switch between an accretion-driven 
emission to a crustal-driven emission is quite fast (Campana et al. 1998; Brown, Bildsten \& Rutledge 1998; Colpi et al. 2001).

When J1023 is in quiescence a radio pulsar has been detected. In this case the X--ray luminosity is lower, $\sim 10^{32}$ erg s$^{-1}$ 
(0.3--10 keV; Archibald et al. 2010; Bogdanov et al. 2011; Li et al. 2014) and pulsations are detected in the radio and X--ray bands. 
The rms pulsed fraction in X--rays is $11\pm2\%$ ($1\,\sigma$ confidence level; Archibald et al. 2011). 
The pulsar X--ray spectrum can be described by a neutron star atmosphere plus power law model. 
This is not directly evident from XMM-Newton (Archibald et al. 2010) and Chandra (Bogdanov et al. 2011) spectra alone, but, when 
fitted together, this spectral model provides a much better description of the data ($5.4\,\sigma$ based on an F-test). 
This model is typical of radio pulsars with a thermal component coming from the polar cap and a power law of magnetospheric origin.
 
The flaring state in this description stands apart. It can be hardly related to an increase of the mass inflow rate since pulsations 
are not even detected. It could instead be ascribed to flaring activity in the disc, possibly related to some magnetic 
reconnection (Zurita, Casares \& Shahbaz 2003) or to some plasmoid ejections in a jet. 
We do not consider this state any further.

\subsection{The spectral model}

Distinguished among LMXBs, J1023 is characterised by three different states when its X--ray luminosity drops below $\sim 10^{34}$ 
erg s$^{-1}$, {apart form the flaring state}.  
According to our picture, in the high state the X--ray spectrum could be described by a composite model made 
by a power law (non-thermal component), an advection dominated disc with free temperature and inner radius (this is achieved 
by using the {\tt DISKPBB} model within XSPEC with $p=0.5$), and a heated polar cap emission modelled with {\tt NSATMOS} 
(Heinke et al. 2006) within XSPEC with a free radius (smaller than the neutron star radius) and temperature. 
The low state is more complex. According to our picture in the low state there should coexist an enshrouded radio pulsar and an outer disc. 
To model the X--ray spectrum in this state we should include a power law non-thermal component (describing either the activity in the 
outermost parts of the disc or, more probably, the shock interaction of the radio pulsar wind with the incoming matter, in this last case 
we would predict $\Gamma\sim 2$, as due to synchrotron radiation), a disc component with an inner radius much larger than the light 
cylinder radius and a radio pulsar emission made by a thermal component and a magnetospheric one ({\tt NSATMOS+POW}). 
We require that the disc temperature and inner disc radius are
such that they lie on the same $T(r)\propto r^{-p}$ curve in the high and low states, i.e. the high state will have a higher temperature 
and a smaller inner disc radius, whereas the low state will have a lower temperature and a larger inner disc radius lying on the 
same $T(r)\propto r^{-p}$ curve. The size of the polar cap is held fixed among the high and low states, even if the temperature is 
allowed to vary.
The magnetospheric pulsar component is free.
Finally, in our picture the quiescent state is completely determined by the low state spectrum with only the pulsar components 
(thermal and non-thermal) being present, but fixed in all their parameters to those of the low state.
All spectra will be corrected for absorption using the same absorbing column density.
See Fig. \ref{fig1} for a picture of our physical scenario.

\section{Data analysis}

We consider three sets of recent observations of J1023 with XMM-Newton. These are long observations carried out during 2013-2014. 
The first two sets (Nov. 2013 and Jun. 2014) were discussed in Bogdanov et al. (2015) and Archibald et al. (2015). 
The last set comprising 4 ToO observations (Nov. 2014) is described in Jaodand et al. (2016, submitted), see Table \ref{obslog} 
for an account of them.

Data reduction was performed with the XMM-Newton Science Analysis Software (SAS) version xmmsas\_20131209\_1901-13.0.0 
and the latest calibration files. We considered only EPIC data. Data were reprocessed with {\tt emproc} and {\tt epproc} locally. 
Data were grade filtered using pattern 0--12 (0--4) for MOS (pn) data, and {\tt FLAG==0} and \#XMMEA\_EM(P) options.
MOS data were all acquired with the thin filter and in small window mode; pn data with the thin filter and in timing (fast-timing) 
mode. Proton flares affected only the latest part of the Jun. 2014 and one of the Nov. 2014 observations, and were filtered out. 
The source events from the pn were extracted using a 7 pixel region centred on source. The MOS events were extracted from a 
870 pixel circular region centred on source. Background events were extracted from similar regions close to the source and free of 
sources. RGS data were not considered in the following, containing less than $10\%$ of MOS counts. 
No narrow emission or absorption lines were detected. We included a zero-width iron line with an energy free to vary within the
6.40-6.93 keV range and derive a $90\%$ upper limit on the equivalent width of $<23$, $<86$, and $<115$ eV, for the high, low, 
and quiescent states, respectively.

\begin{figure}[!ht]
\centerline{
\includegraphics[width=0.5\textwidth]{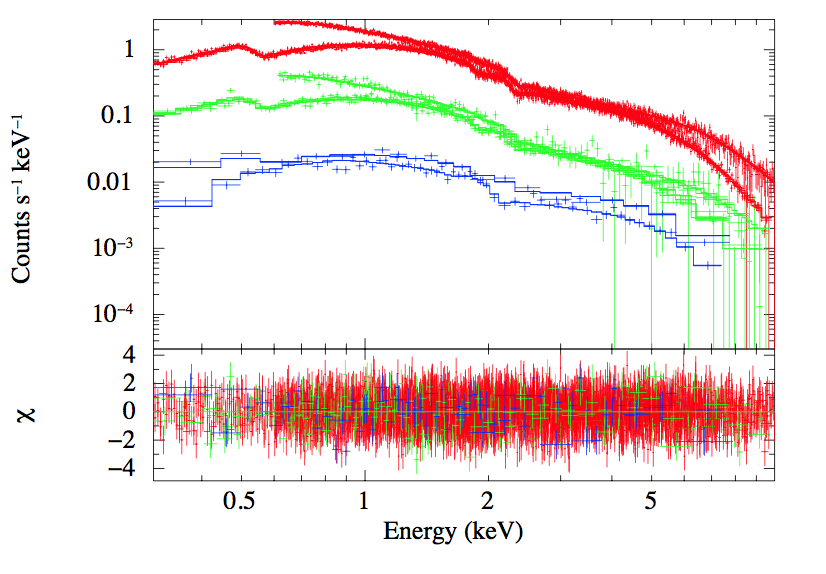}}
\caption{XMM-Newton and Chandra spectra. Upper spectra (in red) refer to the pn (0.6--10 keV) and MOS (0.3--10 keV) high state spectra.
Middle spectra (in green) refer to pn and MOS low state spectra. The lowest spectra (in blue)  refer
to the quiescent state (among the two curves the highest refers to XMM-Newton data and the lowest to Chandra data). 
The fitting model is described in the text.}
\label{fig2}
\end{figure}

\begin{table*}[!htb]
\caption{J1023 X--ray spectral fits.}
\begin{center}
\begin{tabular}{ccccc}
\hline
\hline 
Parameter                                    & High                                  & Low                         & Quiescence\\
                                                       & state                                  & state                        &                      \\   
\hline
$N_H$ ($10^{20}$ cm$^{-2}$)& $5.25^{+0.63}_{-0.60}$ & tied  all                    & tied     all      \\
Power law $\Gamma_D$         &$1.54\pm0.02$                 &$2.00\pm0.03$       & --                  \\
Power law $N_D$ ($10^{-4}$)&$16.7\pm0.5$                   & $2.9\pm0.7$           & --                  \\
Disc $T$ (eV)                              &$138^{+21}_{-17}$         & $<43$                      & --                  \\
Disc norm. $N_d$                      &$103^{+109}_{-55}$       & $>1.1\times 10^4$ & --                   \\
NS atmos. $T$  (eV)                  &$155^{+16}_{-14}$         & $75^{+8}_{-7}$      & tied Low  State   \\
NS atmos. Em. Radius (km)     &$2.6^{+0.6}_{-0.5}$         & tied  all                   & tied all\\
NS Power law $\Gamma_P$   & --                                        &$0.97\pm0.07$      & tied Low State\\
NS Power law $N_P$ ($10^{-5}$)& --                                  &$3.3\pm0.3$           & tied Low State\\
\hline
\hline
\end{tabular}
\end{center}
\noindent $^1$ Errors are $90\%$ confidence level for one parameter of interest. The fit provides a $\chi^2_{\rm red}=1.032$ 
for 3885 degrees of freedom, after the addition of a systematic error of $2\%$. \\
\noindent $^2$ The effective neutron star temperatures were not corrected for the gravitational redshift.\\
\noindent $^3$ Radii were computed using a 1.37 kpc source distance.\\
\label{spec}
\end{table*}

For each observation, we combined the 0.3--10~keV background-subtracted and exposure-corrected light curves of the three CCD 
cameras (all binned at a time resolution of 10~s) using the {\tt epiclccorr} tool in the SAS and the FTOOLS package.
As already reported in previous studies the total light curves are characterised by a low rate mode, a high rate mode, as well as 
random flares. The count rate ranges for the high and low modes are remarkably stable over all observations 
(see also Bogdanov et al. 2015; Archibald et al. 2015; Jaodand et al. 2016). For this reason, we decided to generate good time
intervals by adopting the same thresholds for the count rate ranges as in Bogdanov et al. (2015) to disentangle the different 
modes. With these curves we identified low and high (and flaring) intervals for J1023, following closely Bogdanov et al. (2015) 
prescriptions: good time intervals (GTI) were generated for the low mode requiring the overall count rate to lie in the 
0.0--2.1 c s$^{-1}$ interval and for the high mode in the 4.1--11 c s$^{-1}$. 
Spectra were extracted using these GTIs for the two modes accordingly. Ancillary response files (arf) were generated 
for each spectrum using {\tt arfgen} and redistribution matrices were generated for each observation using {\tt rmfgen}.
MOS1 and MOS2 data and responses were summed for each observation. Data concerning the Nov. 2014 campaign, 
consisting of 4 observations, were summed and response files weighted along the total counts number.
MOS spectra were rebinned to have 100 counts per spectral bin, pn data to have 200 counts due to the higher background.
MOS data were fitted in the 0.3--10 keV energy range, pn data in the 0.6--10 keV due to the larger calibration uncertainties 
in timing mode.

Quiescent data were taken from a 34 ks XMM-Newton observation (Archibald et al. 2010) and a 83 ks Chandra observation 
(Bogdanov et al. 2011). XMM-Newton data were analysed as above, without considering however pn data (operated in timing mode) 
due to the faintness of the source. No proton flare episode is present in the data. MOS1 and MOS2 data were collected with the thin filter
and were summed up. Chandra data were acquired in standard Timed Exposure (TE) mode with the ACIS-S detector.
Data were first reprocessed using the chandra\_repro tool and then analysed using CIAO 4.6 (CALDB 4.5.9).
Photons were extracted from a $5''$ circular region and the background from an annular region free of sources. All the spectra were
rebinned to have 50 counts per spectral bin. Data were retained within the 0.3--8 keV energy range.

\section{Spectral fitting}

We took the existing XMM-Newton data on J1023 (360 ks) and separate them into  a (pure) high state (210 ks), a low 
state (70 ks), and a quiescent radio pulsar  state (34 ks XMM-Newton plus 83 ks Chandra). We fit the corresponding X--ray spectra
with a composite model.

As described above , the high state is modelled by a power law, a radiatively-inefficient disc with free temperature and inner radius 
(this is achieved by using the {\tt DISKPBB} model within XSPEC with $p=0.5$), and a heated polar cap emission modelled with 
{\tt NSATMOS} with a free radius (smaller than the NS radius) and temperature.
 
A different power law, the same radiatively-inefficient disc with free temperature (and radius computed accordingly), together with a 
heated polar cap emission at the same emission radius plus a free power law accounting for the magnetospheric emission was 
adopted for the low state modelling.

The quiescent radio pulsar emission has been fitted with exactly the same neutron star atmosphere component and exactly  the same 
magnetospheric component as the low state, with no additional parameters.
 
All spectra were corrected for absorption using the same absorbing column density\footnote{We tested that the column densities of the 
three states are consistent within the errors, being the one relative to the pulsar quiescent state the lowest.} ({\tt TBABS}, Wilms et al. 2000,
using {\tt VERN} cross sections and {\tt wilm} abundance pattern). 
The same normalisation constant was adopted for all the low state and quiescent MOS spectra (fixed to one).
The other MOS constants were all within $4\%$ of the fixed value. Constants for the pn were all $10\%$ lower 
and  for Chandra ACIS-S $3\%$ higher.
This indicate that the spectra taken in the high and low state over more than one year are really stable.
Given the huge number of counts heavily binned, we assigned a $2\%$ systematic uncertainty to each spectral
channel (added in quadrature), as recommended by Smith et al. (2015),\footnote{See
\url{http://xmm2.esac.esa.int/docs/documents/CAL-TN-0018.pdf}}
The fit worked very well with $\chi^2_{\rm red}=1.032$ for 3885 degrees of freedom (see Table \ref{spec} and Fig. \ref{fig2}),
with a null hypothesis probability of $7.9\%$. In case of no systematic error we have $\chi^2_{\rm red}=1.135$.

We find that the disc component is highly significant in the high state, with an F-test probability of $10^{-9}$ ($6.1\,\sigma$).
On the contrary, the disc component vanishes in the low state.  
The absolute values of the radii are difficult to work out, given the known problems in 
associating the inner disc region to the model normalisation (Kubota et al. 1998), easily involving uncertainties by a factor of $\gsim 2$. 
Taken at face value and with a system inclination of $42^{\rm o}$ (Archibald et al. 2009, 2013), one gets for the high state an 
inner disc radius $r_{\rm high}=21^{+9}_{-7}$ km, very close to the corotation radius of 24 km and fully consistent, given the 
uncertainties, to lie within the corotation radius and the light cylinder radius (i.e. in the propeller state).
The temperature of the disc in the low state cannot be determined but only an upper limit can be set (consistent with our predictions).
This in turn provides a lower limit to the inner disc radius of $\gsim 210$ km. This is well outside the light cylinder radius.

\section{Timing analysis}

Following Archibald et al. (2015), we adopted a definition of the pulsed flux in terms of root-mean-squared (rms) modulation. 
For a signal $f(\phi)$ (with $\phi$) the phase angle the rms modulation is defined as:
$$
F_{\rm rms}=\int (f(\phi) - \bar{f})^2 \, d\phi
$$
where $\bar{f}$ is the mean of the signal.
With this definition Archibald et al. (2015) estimated an rms pulsed fraction of $8.13\pm0.14\%$ in the high state (0.3--10 keV), 
a $95\%$ confidence level upper limit of $<2.8\%$ in the low state (0.3--10 keV), and $11\pm2\%$ in the quiescent radio 
pulsar state in the 0.3--2.5 keV energy range and with a non-detection in the 0.3--10 keV (Archibald et al. 2015). 

Based on our picture, in the low state we have a heavily enshrouded pulsar. The pulsed signal of the pulsar can in principle be
detected in the low state. We rescale the observed pulsed fraction in the quiescent state to the higher total flux in the low state.
We can predict to detect X--ray pulsations in the low state at a level of $\sim 2\%$ in the soft energy band (0.3--2.5 keV).

We searched the two longest uninterrupted XMM-Newton observations (0720030101 and 0742610101) for this pulsation. 
We first isolated the low state as described above. After barycentring the data, we corrected for the 4.75 hr orbital modulation 
and restricted the energy range of the pn data to 0.3--2.5 keV. We then searched for pulsation at the known period,
determined from the analysis of the corresponding high state (and in full agreement with published results, Bogdanov et al. 2014; 
Archibald et al. 2015). We do not detect any pulsed signal and derive $90\%$ confidence level upper limits on the fractional rms 
pulsed emission of $<3.1\%$ and $<3.6\%$ for the two observations, respectively. These were derived fitting the pulse profile with 
two harmonics (as the radio pulsar pulsed profile) and subtracting for the background. These limits, even if tight, still leave open 
the possibility that a radio pulsar-like signal is present in the low state data.

\begin{table*}[!htb]
\caption{0.3--10 keV fluxes of the different components of J1023.}
\begin{center}
\begin{tabular}{ccccc}
\hline
\hline
Parameter     					       & Total/Spectral components& High state                                     &  Low  state                                        & Quiescence\\
							       &					   &						      &						             &	    \\   
\hline
\multirow{5}{*}{Absorbed flux  (\ergscm)}& Total                       	    	   &$1.4\times10^{-11}$                    & $2.0\times10^{-12}$                     	 & $5.6\times10^{-13}$ \\
							       & Power law             		   &$1.2\times10^{-11}$ ($89.8\%$)& $1.4\times10^{-12}$ ($71.7\%$)&  -- \\
							       & Ineffic. disc            		   &$4.8\times10^{-13}$ ($3.4\%$)  & $<0.1 \times10^{-13}$ (--)            & -- \\
							       & NS thermal           		   &$9.5\times10^{-13}$ ($6.8\%$)  & $0.4\times10^{-13}$ ($1.9\%$) & $0.4\times10^{-13}$ ($6.5\%$)\\
							       & Magn. Power law		   &         --                                      & $5.2\times10^{-13}$ ($26.4\%$)&$5.2\times10^{-13}$ ($93.5\%$)\\
\hline
Unabsorbed  flux (\ergscm)   		       & Total				   &$1.5\times10^{-11}$                    & $2.2\times10^{-12}$                        & $5.9\times10^{-13}$ \\
\hline
Luminosity (\ergs)         			       & Total				    &$3.4\times 10^{33}$                    &$5.0\times 10^{32}$                         & $1.3\times 10^{32}$   \\
\hline
\hline
\end{tabular}
\end{center}
\noindent Fluxes refer to the mean of the MOS detectors. The calibration constant is 0.92 for the pn and 1.02 for Chandra/ACIS-S.\\
\noindent The fractional contribution of each component to the total absorbed flux is indicated in parentheses.\\
\noindent The ratio between the high and low state fluxes is 7.1, and the ratio between the low state and quiescent fluxes is 3.5.\\
\label{flux}
\end{table*}

\begin{figure}[!ht]
\centerline{
\includegraphics[width=0.5\textwidth]{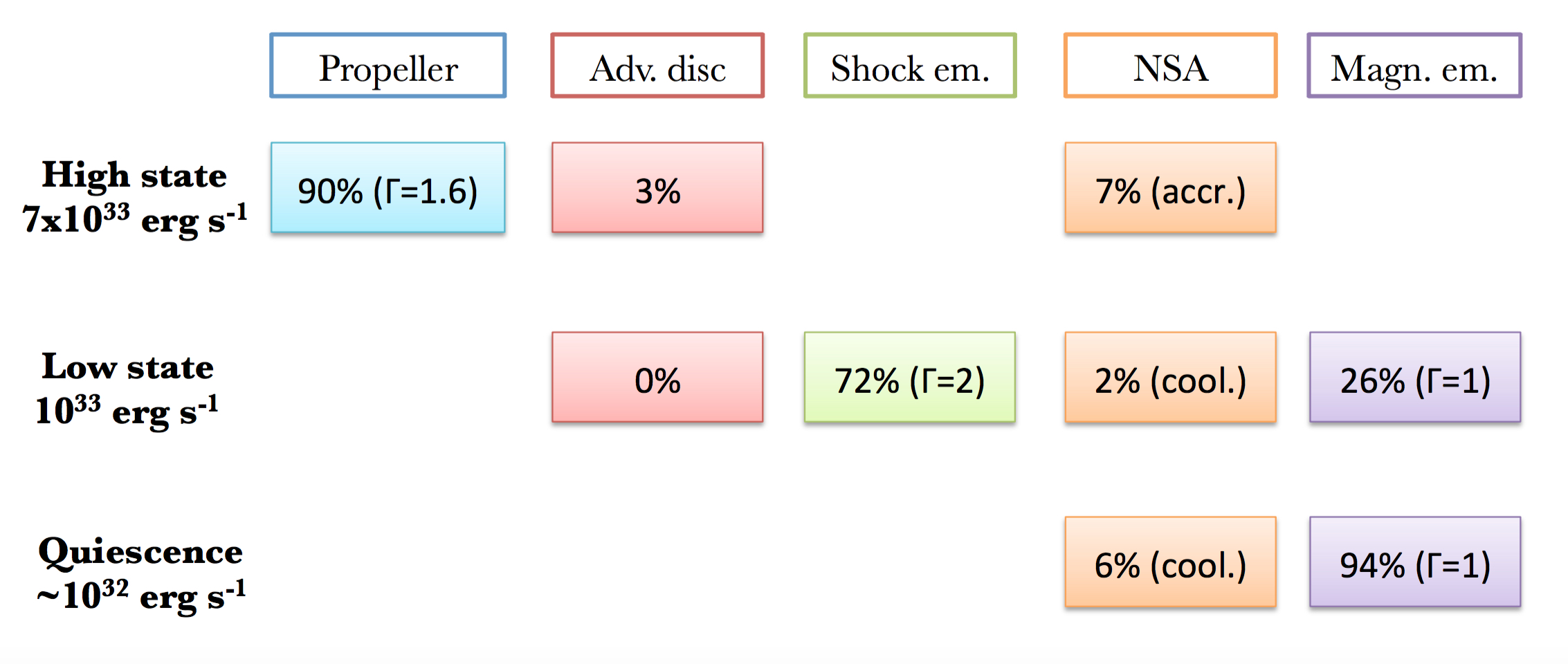}}
\caption{Schematic view of the spectral decomposition proposed in this article. Note that the quiescent emission increases by a factor of 
$\sim 5$ if the power law component extends up to 100 keV.
}
\label{flow}
\end{figure}

\section{Discussion and Conclusions}

Transitional pulsars, and J1023 and XSS  J12270--4859 in particular, show rapid ($\lsim 10$ s) transitions from a high X--ray luminosity state
during which pulsations are observed (rms pulsed fraction $\sim 8\%$) to a low state a factor of $\sim 7$ dimmer and showing no signs 
of pulsations (rms pulsed fraction $<2.4\%$; Bogdanov et al. 2014; Archibald et al. 2015; De Martino et al. 2013; Papitto et al. 2015). 
Similar transitions were also observed in the optical band in J1023, possibly indicating a slightly longer transition time $\sim 20$ s 
(Shahbaz et al. 2015).These state transitions are hardly explained by current modelling of these sources.  

Transitional pulsars and IGR J18245--2452 in particular (Papitto et al. 2013) have shown the possibility to alternate episodes of accretion on the 
neutron star surface to radio pulsar activity on a timescale of less than a week. In this paper we put forward a physical scenario to account for 
rapid transitions among high and low states on a much faster timescale.
We postulate that transitions among the propeller regime and the radio pulsar regime are the main driver for the high and low X--ray 
luminosity states. This implies that J1023 can push matter outside the light cylinder and this can get back close to the corotation 
radius on a 10 s timescale. The dynamical timescale or free fall timescale at the light cylinder radius are much shorter than 10 s 
(of the order of milliseconds), whereas the disc viscous timescale is instead longer (hundreds of seconds). This indicates that 
the disc replenishment should occur on a free-fall-like timescale. Even if we do not investigate the physical mechanism for alternating 
the high and low states, we developed a spectral model to account for the observed spectra as well as the quiescent (radio pulsar) 
state. This model is clearly ad hoc and cannot be proven to be unique, but it provides a satisfactorily description of the observed characteristics.

In the high X--ray state J1023 is in the propeller regime. The X--ray spectrum consists of a non-thermal component, which accounts for the 
majority of the flux. This can come from a hot corona above the disc or (better) from the propeller shocking the inflowing matter. Indeed the 
average Fermi/LAT spectrum is described by a power law with index $1.8\pm0.2$, with a cutoff at an energy of $\sim 2.3$ GeV 
(Takata et al. 2014), consistent with our photon index. A propeller model to account for the main characteristics of the high state has been 
developed by Papitto \& Torres (2015). In their model the inflowing matter is propelled outwards by the rapidly rotating neutron star 
magnetosphere. Electrons can be accelerated to energies of a few GeV at the turbulent disc-magnetosphere boundary. 
Synchrotron and self-synchrotron Compton emission is able to account for the observed X--ray and GeV emission, respectively 
(Papitto \& Torres 2015). This model applies tout-court to our description of the high state.
Additional ingredients to model the high state X--ray spectrum are a radiatively inefficient accretion disc, able to push the inner boundary 
close to the corotation radius but not emitting too much luminosity and some matter leaking through the neutron star magnetosphere to 
account for the pulsed emission. The inner disc radius derived from spectral modelling is consistent to lie close to the corotation radius.
The fraction of the flux emitted from the neutron star component is $7\%$. This flux arises from a small  region on the surface and 
can give rise to pulsations. Its incidence on the total flux is consistent with the observed rms pulsed fraction of $8\%$.

In the low state of J1023 a radio pulsar suddenly turns on. The spectral model to fit the low state X--ray spectrum is made of a power 
law with $\Gamma\sim 2$. This can be interpreted as the shock emission coming from the interaction of the relativistic pulsar wind 
with the inflowing matter. The fraction of spin-down luminosity converted into 0.3--10 keV luminosity is $\sim 2\%$, which is in line with 
other radio pulsars converting their spin-down power into X--rays. The accretion disc in this model is pushed outside the light cylinder radius
and our spectral model does not require the presence of a disc in this state, allowing us to set a lower limit to the disc inner edge well 
outside the light cylinder. Since a radio pulsar is active in this state we added two further spectral components. These are motivated by 
X--ray observations of  radio pulsars. One is thermal in origin arising from the neutron star polar cap and the other is of magnetospheric 
origin. These two components were used to fit simultaneously an XMM-Newton and a Chandra spectra taken when J1023 was in 
quiescence with a radio pulsar detected in the radio band. The radius of the thermal component is kept the same for all the three states. 
This nested spectral model can satisfactorily reproduce the X--ray spectra of the three different states (see Table 3 and Fig. \ref{flow}). 
Even if active during the low state a radio pulsar would be hardly detectable due to the large amount of ionised material 
in the surrounding, producing a large free-free absorption (Campana et al. 1998; Burderi et al 2001; Coti Zelati et al. 2014; 
Deller et al. 2015).

The flat-spectrum radio emission can be associated to an outflow from the system, arising from synchrotron emission. 
This can be in form of a jet (Deller et al. 2015) or it can be launched due to the propeller effect  (Papitto \& Torres 2015).
Optical polarisation observations detected a positive signal in $V$ and $R$ and put a strong upper limit in $I$  (Baglio et al. 2016).
In addition, a  weak modulation with the orbital period is revealed. These observations would argue against a structured jet outflow 
and have been interpreted as due to Thomson scattering of the companion star light with electrons in the disc corona or 
in a large scale outflow (Baglio et al. 2016).

The mechanism responsible for the transition from and to the high and low state is not fully clear. It heavily relies on 
inhomogeneities of the accretion flow to reduce the effective mass inflow rate and lead the magnetosphere to expand 
beyond the corotation radius. The transition from the low state to the high state can be understood in terms of build 
up of matter at the outer disc boundary, able to quench on a very short time scale the turned on radio pulsar. 
Further simultaneous observations from NIR to hard X--rays are clearly needed. This will allow us to study correlations and lags,
and to prove the possible reprocessing of radiation due to the outermost regions of the accretion disc.
In particular, correlations and lags will provide us  a unique tool to investigate the changes in the 
disc size in the two flux modes. According to our picture, the inner disc should be disrupted by the pulsar wind in the low 
mode and the effects on the outer disc need to be explored.

{\it Note added in proof}. After acceptance of this paper, we become aware that a similar scenario has been suggested by Linares et al. (2014) 
to explain high/low states (which were called active/passive modes) in the transitional pulsar IGR J18245-2452 in M?28. 

\begin{acknowledgements}
FCZ, AP and DFT acknowledges the International Space Science Institute (ISSI) Bern, which funded and hosted the international 
team "The disk-magnetosphere interaction around transitional millisecond pulsars". FCZ and NR are supported via an NWO Vidi grant. 
NR and DFT acknowledge support via grants AYA2015- 71042-P and SGR 2014Ð1073.
AP acknowledges support via an EU Marie Sk\"odowska-Curie Individual fellowship under contract no. 660657-TMSP-H2020-MSCA-IF-2014.
\end{acknowledgements}


\end{document}